# Real-time unobtrusive sleep monitoring of in-patients with affective disorders: a feasibility study

**Samuel Askjer**[1,2] · **Kim Mathiasen**[3,4] · **Ali Amidi**[1] · **Christine Parsons**[5] · **Nicolai Ladegaard**[2*]

**Abstract**
Sleep and mental health are highly related concepts, and it is an important research and clinical priority to understand their interactions. In-bed sensors using ballistocardiography provide the possibility of unobtrusive measurements of sleep. In this study, we examined the feasibility of ballistocardiography in measuring key aspects of sleep in psychiatric in-patients. Specifically, we examined a sample of patients diagnosed with depression and bipolar disorder. The subjective experiences of the researchers conducting the study are explored and descriptive analyses of patient sleep are subsequently presented. The practicalities of using the ballistocardiography device seem to be favourable. There were no apparent issues regarding data quality or data integrity. Of clinical interest, we found no link between length of stay and reduced time in bed ($b = -0.06$, $SE = 0.03$, $t = -1.76$, $p = .08$). Using ballistocardiography for measurements on in-patients with affective disorders seems to be a feasible approach.

**Keywords** Affective disorders · ballistocardiography · in-patients · sleep

✉ Nicolai Ladegaard
    nicolai.ladegaard@gmail.com

[1]Department of Psychology & Behavioural Sciences, Aarhus University, Aarhus, Denmark
[2]Department of Affective Disorders, Aarhus University Hospital, Aarhus, Denmark
[3]Centre For Digital Psychiatry, Mental Health Services of Southern Denmark, Odense, Denmark
[4]Department of Clinical Medicine, Faculty of Health Sciences, University of Southern Denmark, Odense, Denmark
[5]Interacting Minds Center, Aarhus University, Aarhus, Denmark

## Introduction

### Depression and sleep

Sleep is critical to physical and psychological functioning. Poor sleep is known to increase the risk for mental disorders and many mental disorders are known to negatively influence sleep. Therefore, there seems to be a bidirectional relationship, where poor sleep and mental disorders cause and maintain each other [1]. Sleep disturbances are generally prevalent across various psychiatric populations hinting at a transdiagnostic pattern [1, 2].

Many patients with depression tend to have difficulties with sleep relating to initiation, maintenance, early awakening, and non-restorative sleep. In addition to this, there tends to be a pattern of reduced slow-wave sleep and disinhibition of rapid-eye movement (REM) sleep. Interestingly, most antidepressants partially work by suppressing REM sleep [3].

Effectively monitoring the development of sleep parameters during acute treatment for depression could be of clinical value [4]. The potential value of automatic sleep monitoring is further emphasized by the current subpar solution, where it is common practice to physically check in on psychiatric in-patients during the night. Automatic unobtrusive sleep assessment, along with psychometrically valid measures of sleep, could be used to identify patients with sleep difficulties and to select appropriate treatment options. Such measurements could also provide a live sleep overview, allowing clinicians to monitor when and how in-patients are sleeping.

## Ballistocardiography

Ballistocardiography is an unobtrusive method for recording pulse, breathing and motion from pressure recoils downwards and in the head-to-foot direction in bed [5]. Characteristics of these frequencies can be used to infer multiple attributes about sleep and bed behaviour [6, 7]. Ballistocardiography has relatively good measurement consistency with polysomnography (PSG) and actigraphy [8, 9].

There is emerging work demonstrating the feasibility of ballistocardiography as a sleep quality indicator [10], with one study reporting on bed occupancy, movement and breathing using this technology. There is currently a wealth of commercially available unobtrusive sleep monitor devices marketed for home usage [11, 12]. However, there is a lack of knowledge about the clinical applicability and measurement accuracy of many such devices for usage with in-patients. This study is, to our knowledge, the first to conduct measurements using ballistocardiography on in-patients with affective disorders.

### Aims

The purpose of this study was to assess the feasibility of using unobtrusive sleep sensor measurements on in-patients hospitalized in a psychiatric unit for affective disorders. More specifically, we aimed to evaluate the practical issues around establishing and collecting such measurements. Additionally, we aimed to assess challenges pertaining to data quality and data handling. Finally, we wanted to gauge the face validity of the acquired results by descriptive analyses of insomnia severity, time in bed and sleep efficiency, as well as preliminarily predicting time in bed based on the length of patients' stay.

## Materials and Methods

### Design

This study was designed as a within-group observational feasibility study. Insomnia severity was measured using a self-reported questionnaire. Participants responded at admission, discharge and at one-week intervals during their admission period. Continuous sleep measurements were recorded during the course of their admissions. During measurements, the participants received treatment as usual.

### Procedure

The participating patients responded to questionnaires in their wards and ballistocardiography devices were placed in their beds. Data collection took place over 16 weeks, from the beginning of November 2020 until the end of March 2021.

### Ethics

Approval was granted by the hospital's internal research committee, clinic manager and attending physician. An inquiry regarding external ethics approval was sent to the Central Jutland Regional Committee on Health Research Ethics in accordance with the Danish Committee Act (LBK nr 1083 of 15/09/2017). The project was considered within ongoing hospital practice and therefore not subject to an external ethical review. The study followed the ethical principles in the Declaration of Helsinki as mandated by The World Medical Association for research on human subjects [14].

Inclusion criteria:
- ≥18 years old
- Bipolar affective disorder (except current mania/hypomania or in remission), depressive episode, recurrent depressive disorder (except in remission).
- Able to communicate in Danish

Exclusion criterion:
- Suicidality (acute heightened risk, as rated by hospital routine screening ranging from no risk (1), heightened (2) to acute heightened (3))

### Recruitment

Eligible patients were approached for enrolment in the study whenever devices were available for use. An initial convenience sample (*n* = 24) was obtained through screening relevant incoming admissions. Participants were briefed on the purpose of the study and gave informed consent for the gathering of data. Three participants later withdrew consent and thus the final sample comprised 21 participants.

### Measurements

Emfit QS™ is a contact-free ferroelectret sensor using ballistocardiography principles; this device was used in the present study [15]. The Emfit model used was the 4th generation IP-9360 and L-0656 sensor, with a three-layered polyester film designed to fit under the mattress. The film is equipped with aluminium electrodes and internal voids, which generate voltage in response to pressure changes [10, 16, 17]. The sensor film measures $542 \times 70 \times 1.4$ mm and is connected by cable to the electronic unit that controls the device and connects to Wi-Fi (IEEE 802.11b/g/n − 2.4 GHz networks), and channels 128 Mbit with a sampling rate of 100 Hz. The electronic unit connects via an external power supply to the AC socket (Fig. 1).

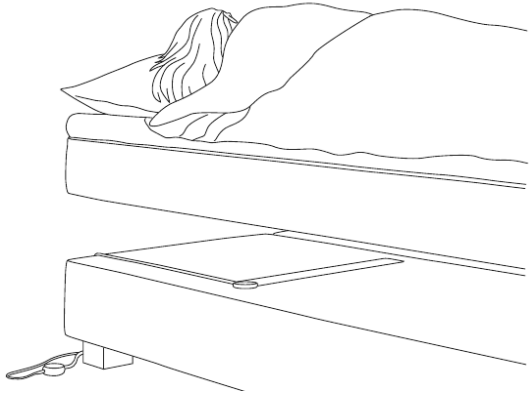

**Fig. 1**
*Sleep sensor setup (Emfit Ltd©)*

The Insomnia Severity Index (ISI) is a seven-point self-report insomnia assessment tool for evaluating difficulties related to sleep onset, sleep maintenance, sleep termination, perceived disturbance severity and impact of insomnia in the patient's life. ISI has demonstrated good internal consistency in clinical populations ($\alpha = .91$) [13].

### Analyses

Qualitative descriptions from the researchers' perspective were based on evaluations from the researchers from conducting the study. Data quality was assessed in terms of availability and completeness, and we also inspected for data integrity issues. Face validity of the results was assessed in terms of descriptive properties, using ISI symptom level change for comparison as well as graphic presentation of time in bed and sleep efficiency.

The potential for predicting time in bed over the course of admission was also considered. This was done using a linear mixed model (LMM), with time as fixed effect and individual participants as random effects (time_in_bed ~ $\text{time}_0$ + [1|participants$_1$] + ε). Missingness was handled by LMM fitted using restricted maximum likelihood estimation from all available data. This allows for handling missing data with limited loss of power. All analyses were performed with Microsoft Excel and Rstudio, with R version 4.0.5 and the packages EnvStats [18] and lme4 [19].

### Results

#### Baseline characteristics

In the final sample (*n* = 21) age ranged from 23 to 72 years and most of the participants were middle-aged ($\bar{x} = 44$, *SD* = 15). The gender balance was thirteen females to eight males. One patient was under clinical evaluation for an affective disorder, 12 patients had bipolar disorder and eight had recurrent depressive disorder according to ICD-10 [20]. The length of stay varied from a week to 53 days, with an average of 21 days (Table 1).

**Table 1.**
*Demographic variables*

| Measure | Freq. | % or range | Mean | Median |
|---------|-------|-----------|------|--------|
| Age | | 23-72 | 44 | 41 |
| 18-29 | 4 | | | |
| 30-49 | 10 | | | |
| 50-64 | 4 | | | |
| >65 | 3 | | | |
| Gender Female | 13 | 61.9 % | | |

|  |  |  |  |  |  |
|---|---|---|---|---|---|
|  | Male | 8 | 38.1 % |  |  |
| Dx. | Bipolar | 12 | 57.1 % |  |  |
|  | Dep. | 8 | 38.1 % |  |  |
|  | Other | 1 | 4.8 % |  |  |
| Length of stay |  |  | 7-53 | 21 | 16 |

## Researchers' perspective and the response from participants

Overall, we had promising evaluations when conducting the feasibility study. Finding the correct position for the device was easy and the device stayed in place with normal bed usage. However, some practical obstacles were encountered e.g., initially we had some technical challenges keeping all devices online due to internal internet (WIFI) security protocols. This was resolved in cooperation with IT services at the hospital. An ongoing challenge at the clinic was abrupt deviations from planned discharges, which caused some data loss. Another challenge was that patients occasionally were transferred to another ward where there was no device installed. Close attention had to be paid to ensure that the device was subsequently moved to the correct ward.

We found that most of the patients that were approached to participate were willing and enthusiastic about having their sleep measured. Many also saw it as an opportunity to learn about their sleep and were interested in their results afterwards. The patients that declined to participate were not expected to offer any explanation. Nonetheless, some patients expressed disinterest or scepticism towards being monitored.

The first participant that withdrew consent got increasingly confused and got increasingly harder to communicate with, this was followed by an inability to fully comprehend what the device represented. The second patient that withdrew had been measured for one night when the device was removed by the patient. According to this patient, it was hard to sleep when "that thing was there". The third patient that withdrew gradually became unmotivated to participate and was discouraged from continuing involvement in the study.

## Acquiring patient data from Emfit

All the 10 devices gave steady bed monitoring data from downloadable csv files. These files could be downloaded on a bed period-to-period basis or by using a multi-day function. We investigated the data integrity of the multi-day function, but found no discrepancies between the period-to-period and the multi-day function. Downloading the data was a fast process and the data files were available for 48 hours in the Emfit cloud solution thereafter. One minor setback was that one cannot download a new multi-day period from the same device in this period. This can be a nuisance when downloading multi-day periods from many patients on the same device.

At one point, a sleep sensor gave invalid results during measurements. This was solved by adjusting the position of the device. There was one instance of a patient removing the device due to believing the measurements were done. This was noticed in the cloud solution and the device was quickly reinstalled. There was also an issue regarding to movement; the activity variable was truncated at 999 and this was occasionally maxed out by some patients. Unstaged (not classified into sleep stages or wakefulness) time in bed was not a major problem, most participants had little or no unstaged time in bed ($\bar{x} = 11\%$). However, two individual participants had a lot of unstaged time in bed (98 & 75%). This was due to lengthy registered bed periods with an average of 35 hr 19 min ($z = 2.87$) and 16 hr 18 min ($z = 2.08$), respectively.

Emfit suspends estimations of wakefulness and sleep when bed periods exceed 14 hours, fall short of 2 hours, or in cases of false reading. Aside from sleep staging and system variables, there were 491 missing values out of 7 810 in total, which amounted to a 6.3% missingness rate. Especially prone to missing values were heart rate variability derived variables.

## Self-reported sleep disturbance, time in bed and sleep efficiency

The mean baseline scores on ISI for the participants were 16.4 ($SD = 7.6$),

corresponding to moderate insomnia. ISI symptom levels decreased throughout the admission periods, but with substantial variations between participants (Fig. 2). Emfit measurements for the entire admission period amounted to a total of 3,354 hours of bed monitoring data. During admission, the total time in bed per nychthemeron decreased slightly (Fig. 3). The LMM showed a non-significant reduction of time in bed over the course of admission ($b$ = -0.06, $SE$ = 0.03, $t$ = -1.76, $p$ = .080), with large variation for the individual participants ($SD$ = 4.02). Thus, the reduction in time in bed amounted to around 3½ min per nychthemeron. Additionally, this model was underpowered with low goodness of fit (log-likelihood = -743). Time in bed was estimated to be awake (14%), asleep (75%) and unstaged (11%). Thus, patient sleep efficiency amounted to 84% of staged time in bed and remained relatively stable over time (Fig. 4).

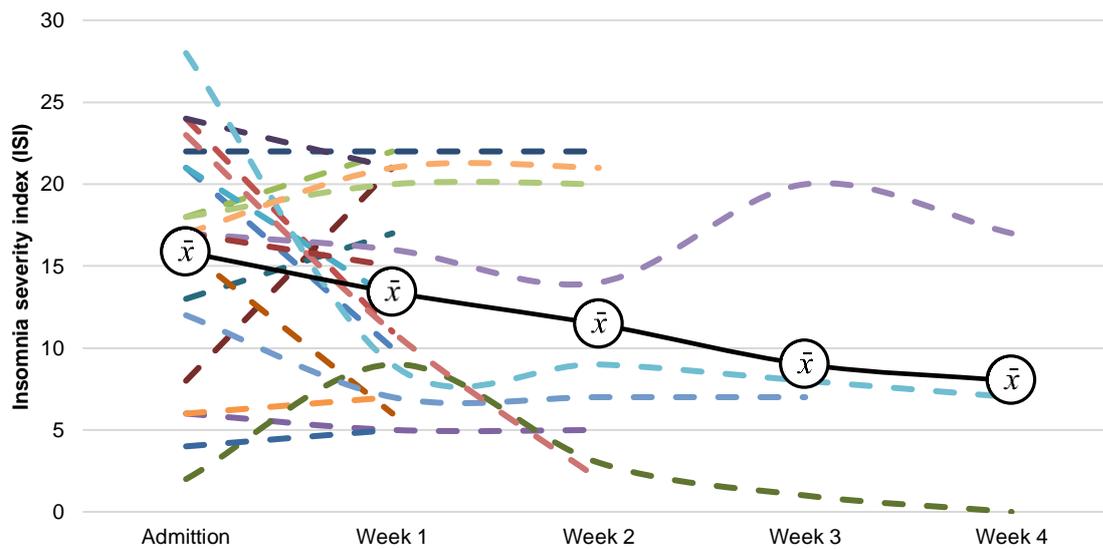

**Fig. 2**
*Sleep disturbance symptom level change on ISI*

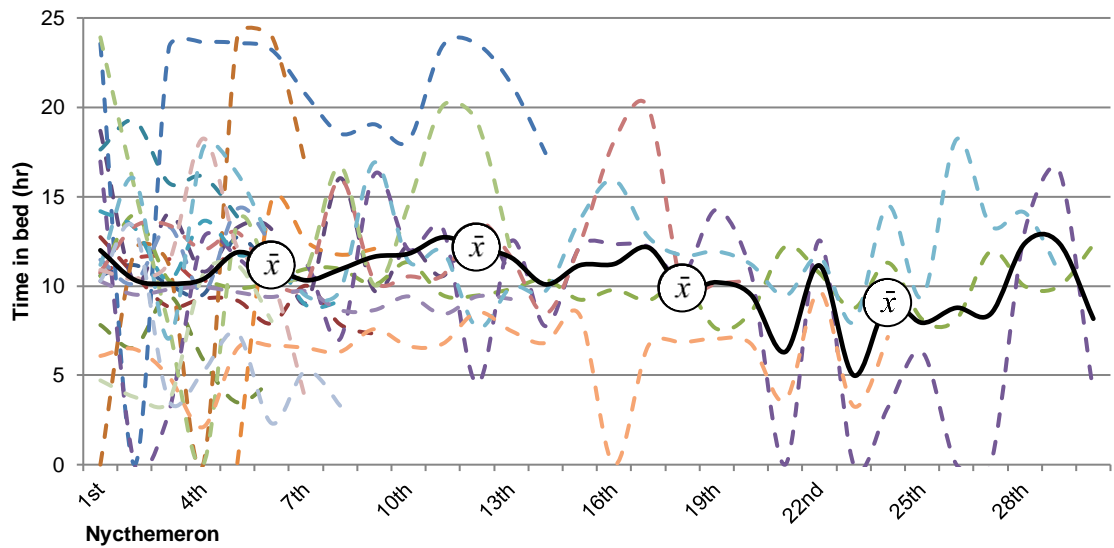

**Fig. 3**

*Time in bed per nychthemeron*

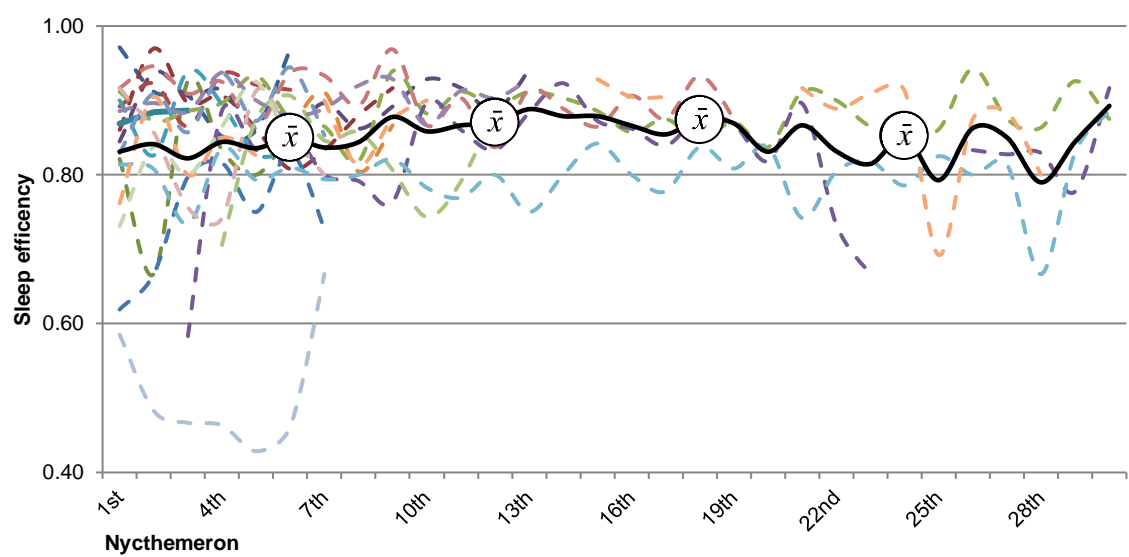

**Fig. 4**
*Sleep efficiency per nychthemeron*

## Discussion

In general, the device used in this study was found to be reliable and appropriate for the setting. The researchers' overall impression was favourable and practical challenges could be managed. Additionally, there were little grounds for data quality and data integrity concerns. There were some challenges with coordinating sleep sensor measurements with patient flow in and out of the clinic, as well as between wards. There were also challenges in regards to minimizing suicide risk.

Most patients were motivated to partake in the study; some patients that declined to participate were either unmotivated, disinterested, or sceptical. There could be reasons to think that bed monitoring would be challenging with patients that have anxiety related to sleep or surveillance apprehension. This assumption is based on our experience with several patients who either declined involvement or withdrew consent with such concerns. The clearest challenge in regards to acquiring data was related to invalid and missing data. Extensive time in bed could lead to more unstaged bed periods and thus causing more missingness.

Reduced time in bed and consequently short sleep duration (<6 hr) has been tied to increased risk of developing depression, but there are also indications that extensive time in bed could increase depression risk [21, 22]. Sleep efficiency is typically seen to improve alongside remission from depressive symptoms [23]. In addition to frequent comorbidity with insomnia, depression and bipolar disorder are often comorbid with hypersomnia, which can lead to increased bed usage [24-26].

### Limitations and future directions

There were several limitations in this study that need to be considered. Sleep efficiency was likely overestimated because excessive time spent awake in bed could be construed by the device as sleeping [27]. The ability of the device to distinguish between wakefulness and sleep has previously been validated, but not in a depressed population [17].

Future research should strive to overcome challenges relating to suicidality innate to using a measurement device with an attached cord. This could be done by fastening the device to the bed in such a way that suicide is unattainable. We made efforts to zip-tie the device to the bed, but no fully satisfactory solution was reached. Another solution would be to use completely wireless sleep sensors e.g., radio frequency identification tags [28, 29], ultra wideband pulse-Doppler radar [30], dual pulse-Doppler radar [31, 32] or frequency-modulated continuous wave Doppler radar [33].

Validation studies using ballistocardiography principles and PSG on an in-patient sample with affective disorders would also be preferable before further exploring the potential of this device. A larger experimental trial for inferential purposes could be appropriate if validation studies generalize to this setting.

## Conclusion

Ballistocardiography seems to present a feasible approach for measuring the sleep of in-patients with affective disorders. Over the course of admission, there was a non-significant reduction in the average time in bed per nychthemeron. In summary, the findings presented in this study could serve as a knowledge base for the further use of unobtrusive sleep sensors on this population and Inform further research on the topic.

## Acknowledgments

We want to acknowledge the contribution of the research and clinic staff at the department of affective disorders at Aarhus University Hospital; research statistician Maria Speed in particular. Lastly, we want to give our thanks to the participants that took the time and effort to answer our questions and allowed for measurements of their sleep.

## Statements and Declarations

No funding was granted for the completion or publication of this study, Emfit units were sponsored by the research committee at Aarhus University Hospital. No conflicts of interest to declare. Nicolai Ladegaard conceptualized and organized the groundwork for the study. Collection of data and analyses was conducted primarily by Samuel Askjer with support from Nicolai Ladegaard. The main body of the manuscript was written by Samuel Askjer with significant contributions from all authors. Furthermore, all authors commented on revisions and approved the final manuscript. Ethical approval was granted by the internal research committee at Aarhus University Hospital. All subjects gave informed consent prior to participation. All subjects gave consent for the publication of relevant anonymized measurements.

## References

1. Freeman D, Sheaves B, Waite F, Harvey AG, Harrison PJ. Sleep disturbance and psychiatric disorders. The Lancet Psychiatry. 2020; https://doi.org/10.1016/S2215-0366(20)30136-X

2. Kelly JM, Strecker RE, Bianchi MT. Recent developments in home sleep-monitoring devices. International Scholarly Research Notices. 2012; https://doi.org/10.5402/2012/768794

3. Sivertsen B, Hysing M, Harvey AG, Petrie KJ. The epidemiology of insomnia and sleep duration across mental and physical health: the SHoT study. Frontiers in Psychology. 2021; https://doi.org/10.3389/fpsyg.2021.662572

4. Riemann D, Krone LB, Wulff K, Nissen C. Sleep, insomnia, and depression. Neuropsychopharmacology. 2020 Jan;45(1):74-89.

5. Bei B, Asarnow LD, Krystal A, Edinger JD, Buysse DJ, Manber R. Treating insomnia in depression: Insomnia related factors predict long-term depression trajectories. Journal of consulting and clinical psychology. 2018; https://doi.org/10.1037/ccp0000282

6. Mack DC, Patrie JT, Suratt PM, Felder RA, Alwan M. Development and preliminary validation of heart rate and breathing rate detection using a passive, ballistocardiography-based sleep monitoring system. IEEE Transactions on Information Technology in Biomedicine. 2008; 10.1109/TITB.2008.2007194

7. Park KS, Hwang SH, Yoon HN, Lee WK. Ballistocardiography for nonintrusive sleep structure estimation. In2014 36th Annual International Conference of the IEEE Engineering in Medicine and Biology Society 2014; https://doi.org/10.1109/EMBC.2014.6944793

8. Pinheiro E, Postolache O, Girão P. Theory and developments in an unobtrusive cardiovascular system representation: ballistocardiography. The open biomedical engineering journal. 2010; https://doi.org/10.2174%2F1874120701004010201

9. Jaworski DJ, Roshan YM, Tae CG, Park EJ. Detection of sleep and wake states based on the combined use of actigraphy and Ballistocardiography. In2019 41st Annual International Conference of the IEEE Engineering in Medicine and Biology Society (EMBC) 2019; https://doi.org/10.1109/EMBC.2019.8857650

10. Piantino J, Luther M, Reynolds C, Lim MM. Emfit Bed Sensor Activity Shows Strong Agreement with

Wrist Actigraphy for the 'Assessment of Sleep in the Home Setting. Nature and Science of Sleep. 2021; https://doi.org/10.2147/NSS.S306317

11. Ramos-Castro J, Mahdavi H, de Cerio DP, García-González MA, Estevez S, Rosell-Ferrer J. Unobtrusive Activity Measurement in Patients with Depression: the H4M Approach. Information and Communication Technologies applied to Mental Health.:18.

12. Hussain Z, Sheng QZ, Zhang WE, Ortiz J, Pouriyeh S. A review of the non-invasive techniques for monitoring different aspects of sleep. arXiv preprint arXiv:2104.12964. 2021 Apr 27; https://doi.org/10.48550/arXiv.2104.12964

13. Morin CM, Belleville G, Bélanger L, Ivers H. The Insomnia Severity Index: psychometric indicators to detect insomnia cases and evaluate treatment response. Sleep. 2011; https://doi.org/10.1093/sleep/34.5.601

14. World Medical Association. World Medical Association Declaration of Helsinki: Ethical Principles for Medical Research Involving Human Subjects. JAMA. 2013; https://doi.org/10.1001/jama.2013.281053

15. Emfit ltd. Home, 2020a.

16. Emfit ltd, Emfit QS: Installation & operating instructions, Vaajakoski, Finland, 2017.

17. Kortelainen JM, Mendez MO, Bianchi AM, Matteucci M, Cerutti S. Sleep staging based on signals acquired through bed sensor. IEEE Transactions on Information Technology in Biomedicine. 2010; https://ieeexplore.ieee.org/document/544769

18. Millard SP, EnvStats-an R. An R package for environmental statistics. New York: Springer-Verlag; 2013.

19. Bates D, Mächler M, Bolker B, Walker S. Fitting linear mixed-effects models using lme4. arXiv preprint arXiv:1406.5823. 2014 Jun 23.

20. ICD-10 Classifications of Mental and Behavioural Disorder: Clinical Descriptions and Disgnostic Guidelines. Geneva. World Health Organisation. 1992.

21. Furihata R, Uchiyama M, Suzuki M, Konno C, Konno M, Takahashi S, Kaneita Y, Ohida T, Akahoshi T, Hashimoto S, Akashiba T. Association of short sleep duration and short time in bed with depression: AJ apanese general population survey. Sleep and Biological Rhythms. 2015; https://doi.org/10.1111/sbr.12096

22. Reynold AM, Bowles ER, Saxena A, Fayad R, Youngstedt SD. Negative effects of time in bed extension: a pilot study. Journal of sleep medicine and disorders. 2014 Aug 28;1(1).

23. Thase ME. Depression and sleep: pathophysiology and treatment. Dialogues in clinical neuroscience. 2022; https://doi.org/10.31887/DCNS.2006.8.2/mthase

24. Dauvilliers Y, Lopez R, Ohayon M, Bayard S. Hypersomnia and depressive symptoms: methodological and clinical aspects. BMC medicine. 2013; https://doi.org/10.1186/1741-7015-11-78

25. Harvey AG. Sleep and circadian rhythms in bipolar disorder: seeking synchrony, harmony, and regulation. American journal of psychiatry. 2008; https://doi.org/10.1176/appi.ajp.2008.08010098

26. Soehner AM, Kaplan KA, Harvey AG. Prevalence and clinical correlates of co-occurring insomnia and hypersomnia symptoms in depression. Journal of affective disorders. 2014; https://doi.org/10.1016/j.jad.2014.05.060

27. Emfit ltd. Features, 2020b.

28. Hu X, Naya K, Li P, Miyazaki T, Wang K. Non-invasive sleep monitoring based on RFID. In2017 IEEE 19th International Conference on e-Health Networking, Applications and Services (Healthcom) 2017; https://doi.org/10.1109/HealthCom.2017.8210832.

29. Sharma P, Kan EC. Sleep scoring with a UHF RFID tag by near field coherent sensing. In2018 IEEE/MTT-S International Microwave Symposium-IMS 2018; https://doi.org/10.1109/MWSYM.2018.8439216

30. Toften S, Pallesen S, Hrozanova M, Moen F, Grønli J. Validation of sleep stage classification using non-contact radar technology and machine learning (Somnofy®). Sleep Medicine. 2020; https://doi.org/10.1016/j.sleep.2020.02.022

31. Tran VP, Al-Jumaily AA. Non-contact Doppler radar based prediction of nocturnal body orientations using deep neural network for chronic heart failure patients. In2017 International Conference on Electrical and Computing Technologies and Applications (ICECTA) 2017; https://doi.org/10.1109/ICECTA.2017.8252020

32. Zaffaroni A, De Chazal P, Heneghan C, Boyle P, Mppm PR, McNicholas WT. SleepMinder: an innovative contact-free device for the estimation of the apnoea-hypopnoea index. In2009 annual international conference of the IEEE engineering in medicine and biology society 2009; https://doi.org/10.1109/IEMBS.2009.5332909

33. Postolache OA, Girão PS, Postolache G. Comparative analysis of two systems for unobtrusive heart signal acquisition and characterization. In2013 35th Annual International Conference of the IEEE Engineering in Medicine and Biology Society (EMBC) 2013; https://doi.org/10.1109/EMBC.2013.6611174